\documentclass[conference]{IEEEtran}

\usepackage[dvips]{graphicx}

\ifCLASSINFOpdf
\else
\fi

\usepackage[tight,footnotesize]{subfigure}
\usepackage{footnote}
\makesavenoteenv{tabular}

\begin{document}

\title{Using the Palomar Transient Factory to Search for Ultra-Long-Period Cepheid Candidates in M31}

\author{
\IEEEauthorblockN{Chien-Hsiu Lee\IEEEauthorrefmark{1}, Chow-Choong Ngeow\IEEEauthorrefmark{1}, Michael Ting-Chang Yang\IEEEauthorrefmark{1}\IEEEauthorrefmark{2}, Wing-Huen Ip\IEEEauthorrefmark{1},\\ Albert Kwok-Hing Kong\IEEEauthorrefmark{3}, Russ R. Laher\IEEEauthorrefmark{4} and Jason Surace\IEEEauthorrefmark{4}}
\IEEEauthorblockA{\IEEEauthorrefmark{1}Graduate Institute of Astronomy, National Central University, Jhongli 32001, Taiwan \\
Email: cngeow@astro.ncu.edu.tw}
\IEEEauthorblockA{\IEEEauthorrefmark{2}Cahill Center for Astrophysics, California Institute of Technology, Pasadena, CA 91125, USA}
\IEEEauthorblockA{\IEEEauthorrefmark{3}Institute of Astronomy and Department of Physics, National Tsing Hua University, Hsinchu 30013, Taiwan}
\IEEEauthorblockA{\IEEEauthorrefmark{4}Spitzer Science Center, California Institute of Technology,  M/S 314-6, Pasadena, CA 91125, U.S.A.}
}

\maketitle

\begin{abstract}

Ultra-long-period Cepheids (ULPCs) are important in distance-scale studies due to their potential for determining distance beyond $\sim100$~Mpc. We performed a comprehensive search for ULPCs in M31, a local benchmark to calibrate the distance ladders. We use data from the Palomar Transient Factory (PTF), which has imaged M31 using a 1.2-m telescope equipped with a $\approx$7.26~deg$^2$ field-of-view (FOV) camera, usually with daily sampling, since the beginning of 2010. The large FOV, together with the regular monitoring, enables us to probe ULPCs in the bulge, disk, and even out to the halo of M31. Using a difference imaging analysis technique, we found and characterized 3 promising ULPC candidates based on their luminosities, amplitudes and Fourier parameters. The mean absolute magnitude for these 3 ULPC candidates, calibrated with latest M31 distance, is $M_R=-6.47$mag. Two out of the 3 ULPC candidates have been reported in literature, however their published periods from Magnier et al. are about half of the periods we found in this work. The third ULPC candidate is a new discovery. We studied 5 other candidates and determined that they are probably Mira-like or ultra-long-period variables, but not ULPCs. 

\end{abstract}


\IEEEpeerreviewmaketitle

\section{Introduction}

The ultra-long-period Cepheids (ULPCs) are defined as classical fundamental mode Cepheids with pulsating period longer that 80~days (\cite{bird09}). They are located at the bright end of the period-luminosity (P-L) relation defined by the classical, shorter period Cepheids, but follow a flatter P-L relation (\cite{bird09}). This suggests that, together with their intrinsic brightness of $M_I\approx -7.86$mag, the ULPCs can be used as standard candle to find distances up to $\approx100$~Mpc and beyond using current and future space-based telescopes (\cite{bird09}). Galaxies with distance determined from ULPCs can then be used to calibrate a number of secondary distance indicators, such as Tully-Fisher relations and type Ia supernovae, that are well within the Hubble flow. \cite{bird09} listed 18 ULPCs from 6 galaxies in the Local and Sculptor Groups, most of them are dwarf or irregular galaxies. \cite{fiorentino12} include 19 additional ULPCs from 4 nearby spiral galaxies. Surprisingly, none of these galaxies include M31 --- the nearest spiral galaxy. A number of investigators has suggested to use M31 as the calibrating, or anchoring, galaxy in distance scale determination (for example, see \cite{vilardell10} and reference therein). As the importance of M31 in distance scale work is increasing, it is necessary to search and calibrate the ULPCs in M31. In this work, we report our finding of ULPC candidates in M31 using the imaging data taken from the Palomar Transient Factory project.

\section{The Palomar Transient Factory}

The Palomar Transient Factory (PTF, \cite{rau09,law09}) is one of the leading survey project in time-domain astronomy. The aim of PTF is to discover transients in the Universe, including various types of supernovae as well as other exotic and rare transient events. On average, PTF is now finding one transient every 20 minutes and one strong variable (with variation greater than 10\%) every 10 minutes. The PTF data can also be used to study other time-varying objects, such as different types of variable stars and asteroids. PTF utilizes the 48-inch Samuel Oschin Telescope (also known as P48) at Palomar Observatory for discovering the transients, and follow-up mainly by the P60 Telescope (for photometric light and color curves) and other large aperture telescopes within the PTF network (for spectroscopic classification, including Palomar 200-inch Telescope and Keck Telescopes). 

\begin{figure*}[!t]
\centering
\includegraphics[width=6.2in]{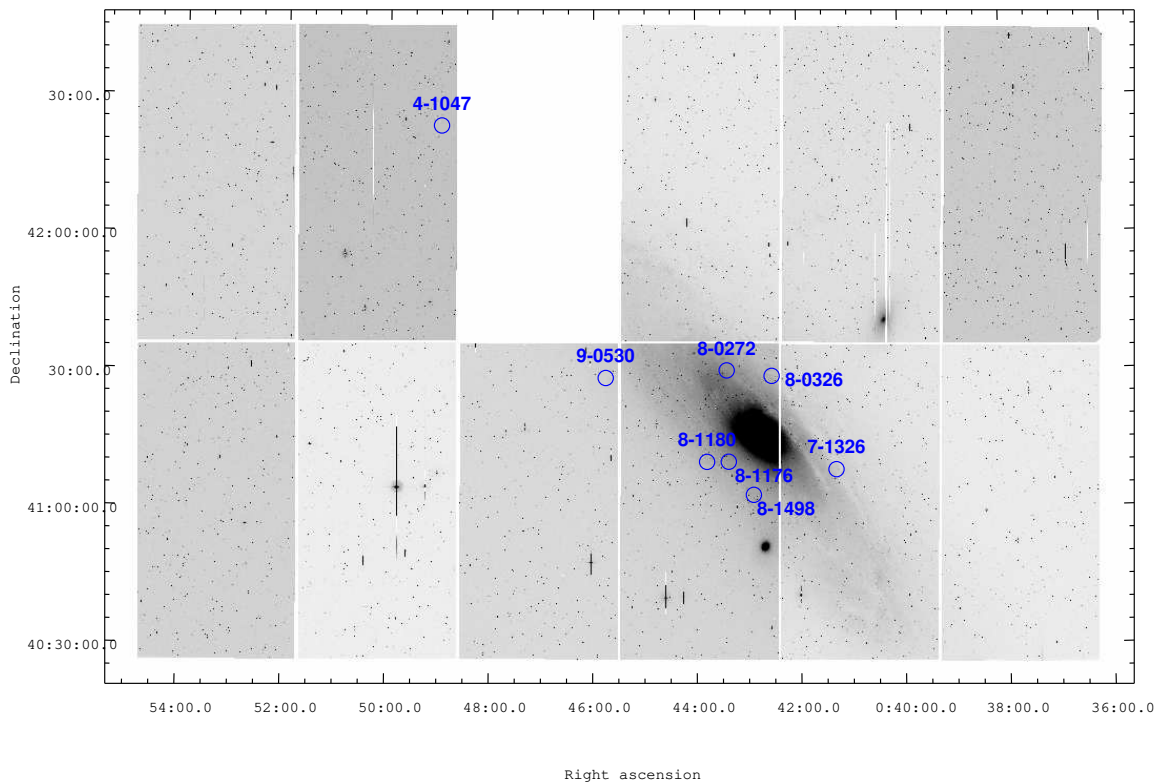}
\caption{Mosaic image for M31 based on PTF data, which also shows the FOV of the P48, as well as the relative position of M31 in PTF field ID~\#4445. Note that one CCD (upper row, third from the left) is inoperable. Locations of the 8 candidates that survived after the cuts (see Section~\ref{sec3}) are also indicated. Note that out of the 8 candidates, only three of them being classified as ULPC candidates (see Section~\ref{sec4}).}
\label{fig_m31}
\end{figure*}

The wide-field P48 Schmidt telescope has been used for the famous Palomar Observatory Sky Survey (POSS) to map out the entire Northern sky with photographic plates. Today, the robotic P48 is equipped with a CFH12K mosaic camera that has been redesigned to fit the telescope. The pixel size of the CCD camera is 1.01~arcsec/pixel, with 11 operable CCD arrays, giving a $\approx$7.26~deg$^2$ field-of-view (FOV) in total (see Figure~\ref{fig_m31}). Most of the PTF observations on P48 telescope are done in Mould $R$-band filter, with occasionally use the $g$ or $H\alpha$ filters. The nightly images taken for P48 telescope are processed with standard image-reduction procedures, which include bias subtraction, flat-fielding, artifact flagging, and astrometric and photometric calibration (\cite{grillmair10,ofek12}). The processed images are archived in a database at the Infrared Processing and Analysis Center (IPAC), which is accessible only by PTF-collaboration members. 

\section{Searching of ULPCs in M31 Using PTF Data}
\label{sec3}

The PTF data is ideal for the search of ULPCs in M31. This is because the FOV of P48 is able to cover the entire M31 in a single exposure, as shown in Figure~\ref{fig_m31}. Furthermore, M31 has been substantially observed by P48 since the beginning of 2009, whenever it is visible, with a time resolution of up to 1 day. Two sets of data are available from PTF data archive, with PTF Field ID~\#4445 and \#100043, respectively. In this work, we specifically downloaded images with PTF Field ID~\#4445 to search for the ULPCs. This particular M31 data set was observed starting on 17 January 2010 and ending on 31 January 2012, with a total of 172 frames. 

\begin{figure}
\centering
\includegraphics[width=2.5in]{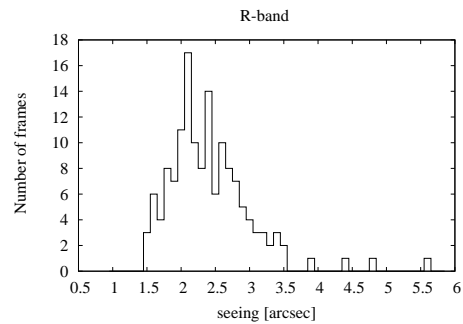}
\caption{Seeing distribution of the P48 images studied in this work. The value of the seeing is given by running {\tt SExtractor} (\cite{bertin96}) on the reduced images, which has been written to the header of the FITS file.}
\label{fig.psf}
\end{figure}

After downloading the reduced images from IPAC, we performed image subtraction following the difference imaging analysis (DIA) proposed by \cite{alard98} for each CCD as follows. First of all, to get rid of the shifts and rotations between images, we registered all the images onto an astrometric reference, which was chosen to be the best-seeing image from all of the available times (the seeing distribution of the images is presented in Figure~\ref{fig.psf}). Then, we re-mapped all the images to this astrometric reference with a second-degree polynomial interpolation along both image horizontal ($x$-axis) and vertical ($y$-axis) dimensions.

\begin{table*}
\renewcommand{\arraystretch}{1.3}
\caption{Surviving number of M31 ULPC candidates after each of the applied cuts (see Section~\ref{sec3}).}
\label{tab1}
\centering
\begin{tabular}{l|ccccccccccc}
\hline
Criterion & \multicolumn{11}{c}{CCD number} \\
\hline
                                 &    0  &     1 &    2  &     4 &     5 &    6  &    7  &    8  &    9  &    10 &    11   \\
\hline
17.5 $\le$ $R$ $\le$ 19.5        &  3059 &  2560 &  3244 &  3070 &  3101 &  2450 &  3792 &  5162 &  2555 &  2341 &  2639   \\
0.5 $\le$ Power$_{AoV}$ $\le$ 0.9 &  1442 &  1494 &   551 &   141 &   178 &    48 &    73 &   155 &  1584 &   552 &   643   \\ 
80 d $\le$ Period $\le$ 300 d    &   490 &    53 &    11 &     3 &     2 &     0 &     4 &    14 &     3 &     2 &     1   \\
Eye inspection                   &     0 &     0 &     0 &     1 &     0 &     0 &     1 &    5  &     1 &     0 &     0   \\
\hline
\end{tabular}
\end{table*}

\begin{figure*}
  \centering
  $\begin{array}{cccc}
  \includegraphics[width=0.22\textwidth]{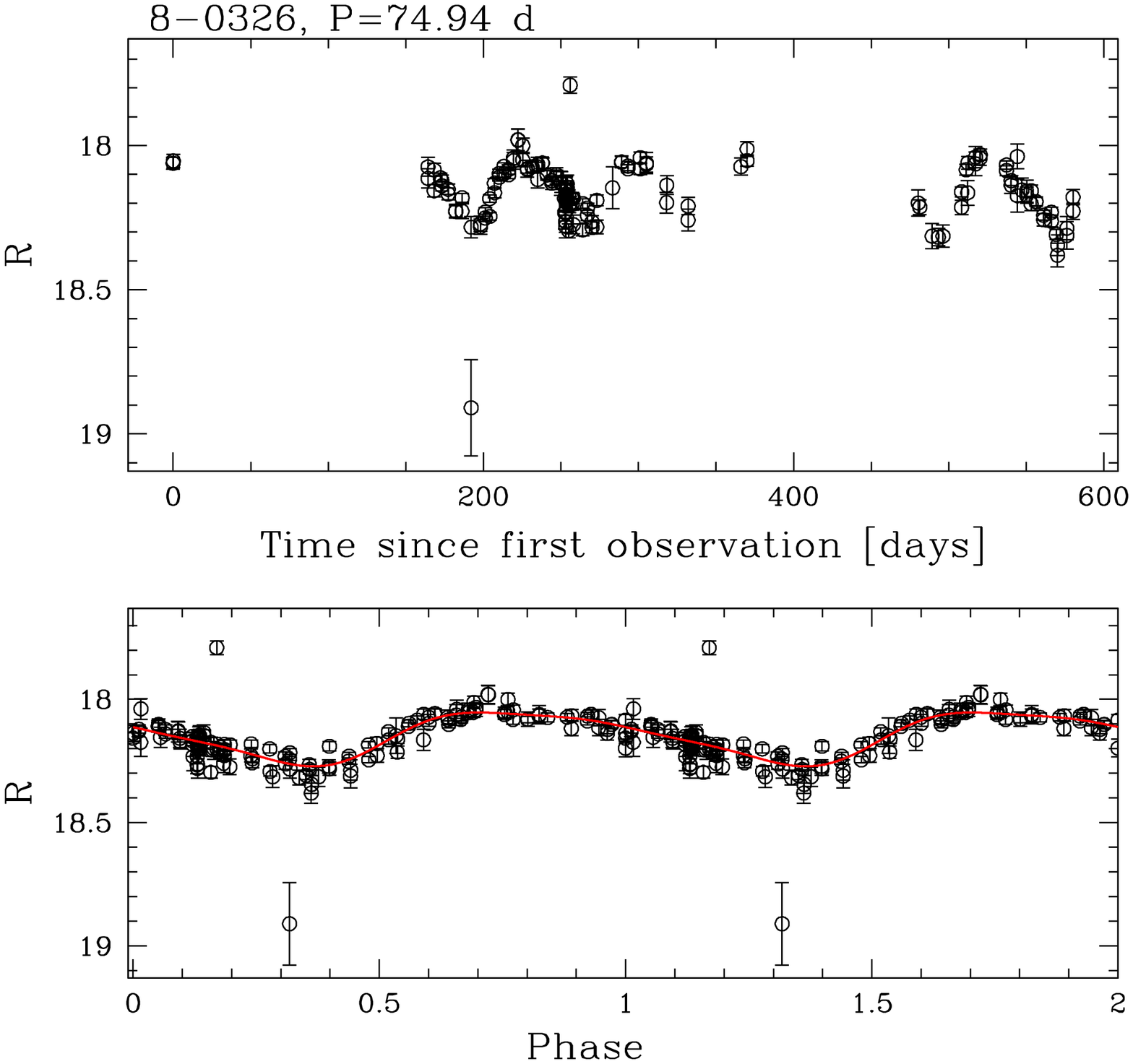} &
  \includegraphics[width=0.22\textwidth]{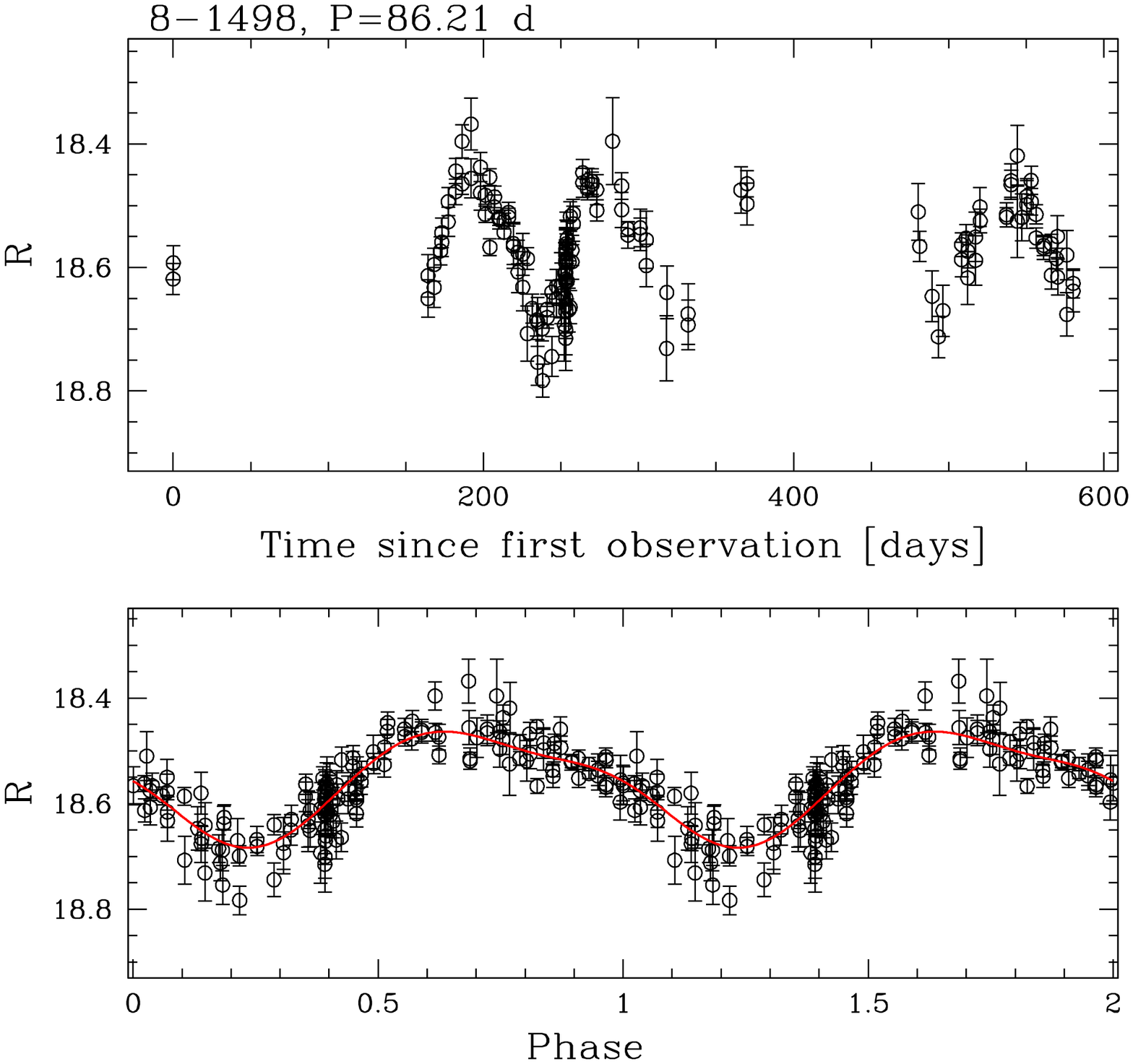} &
  \includegraphics[width=0.22\textwidth]{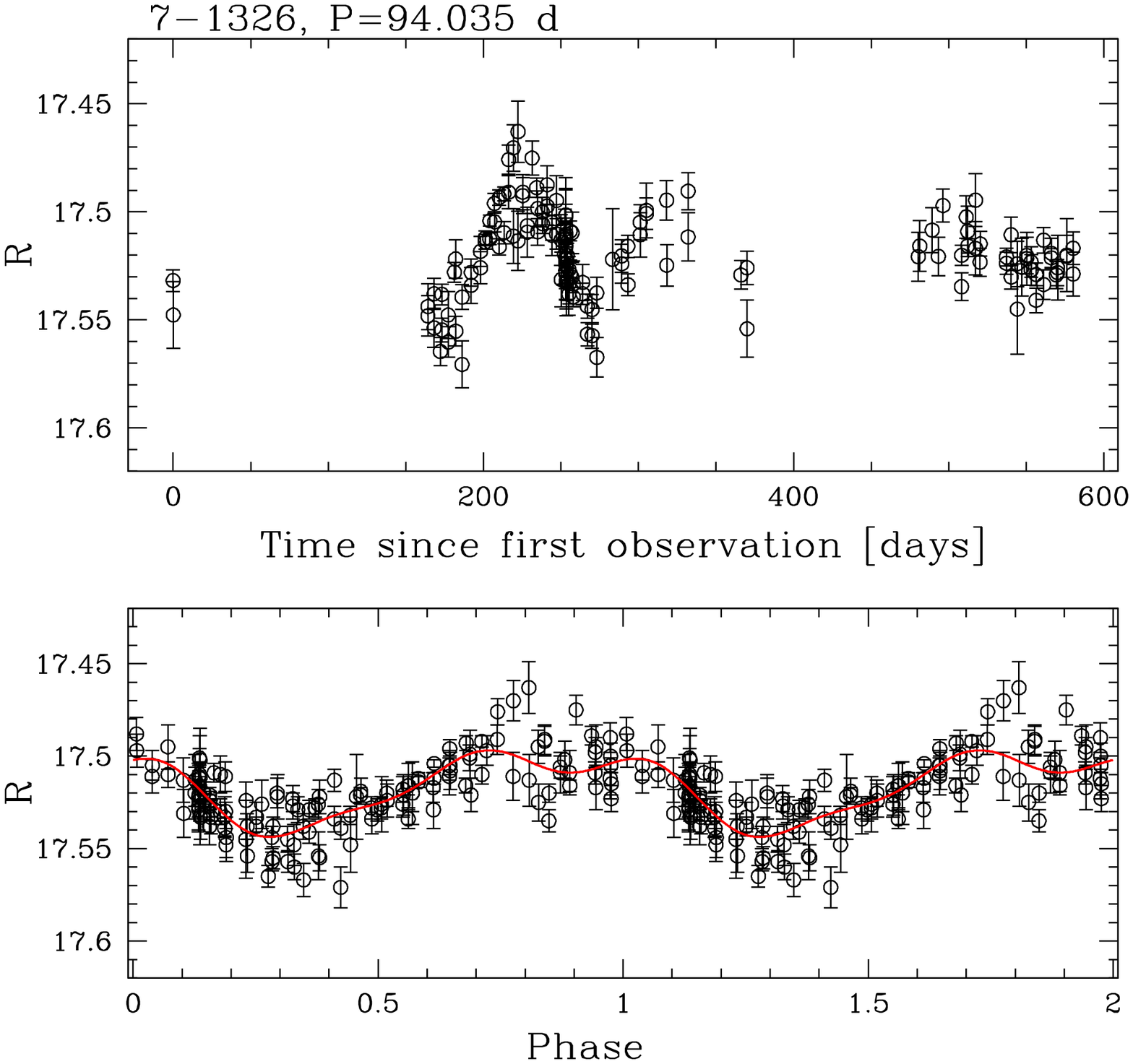} &
  \includegraphics[width=0.22\textwidth]{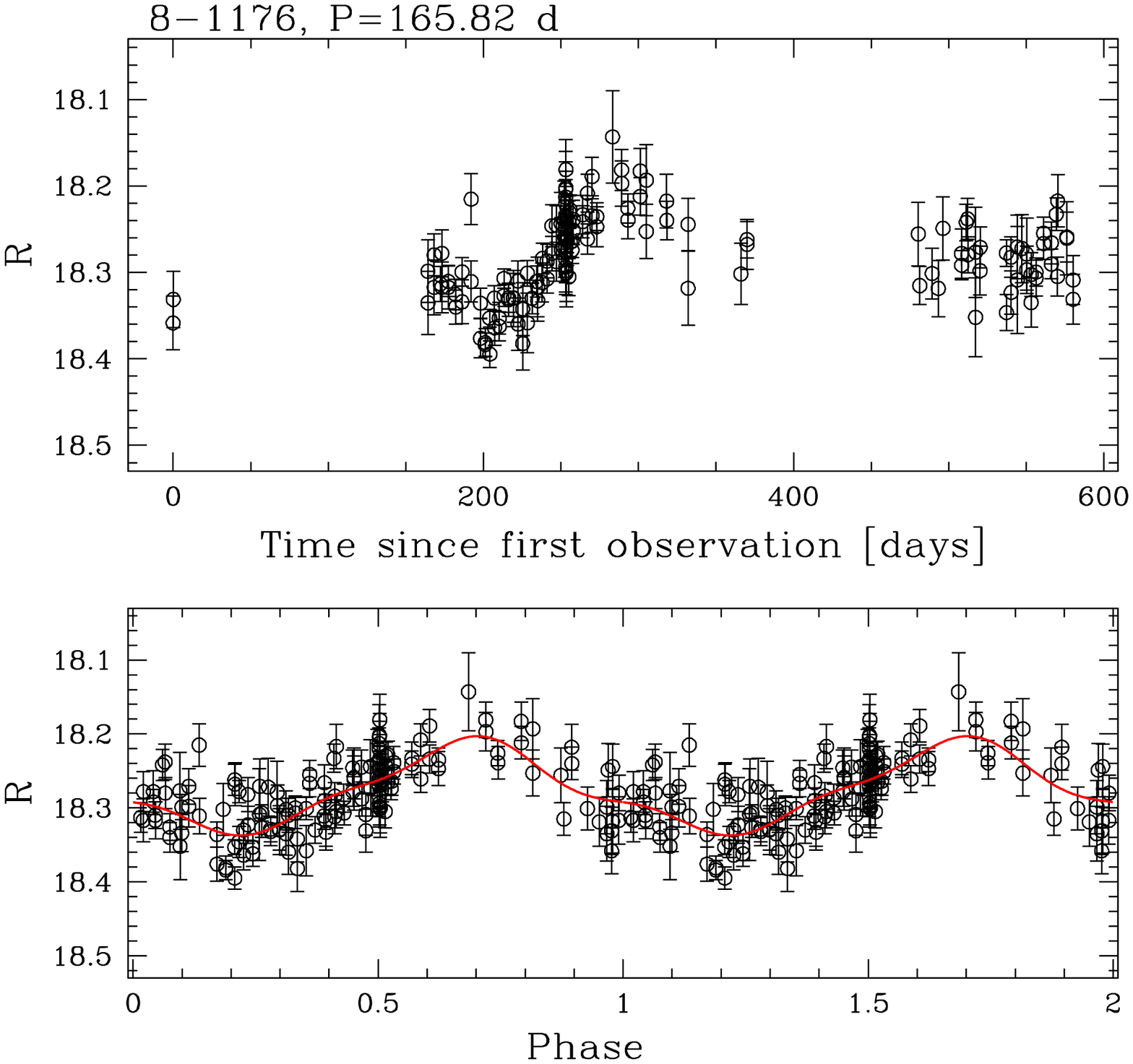} \\
  \includegraphics[width=0.22\textwidth]{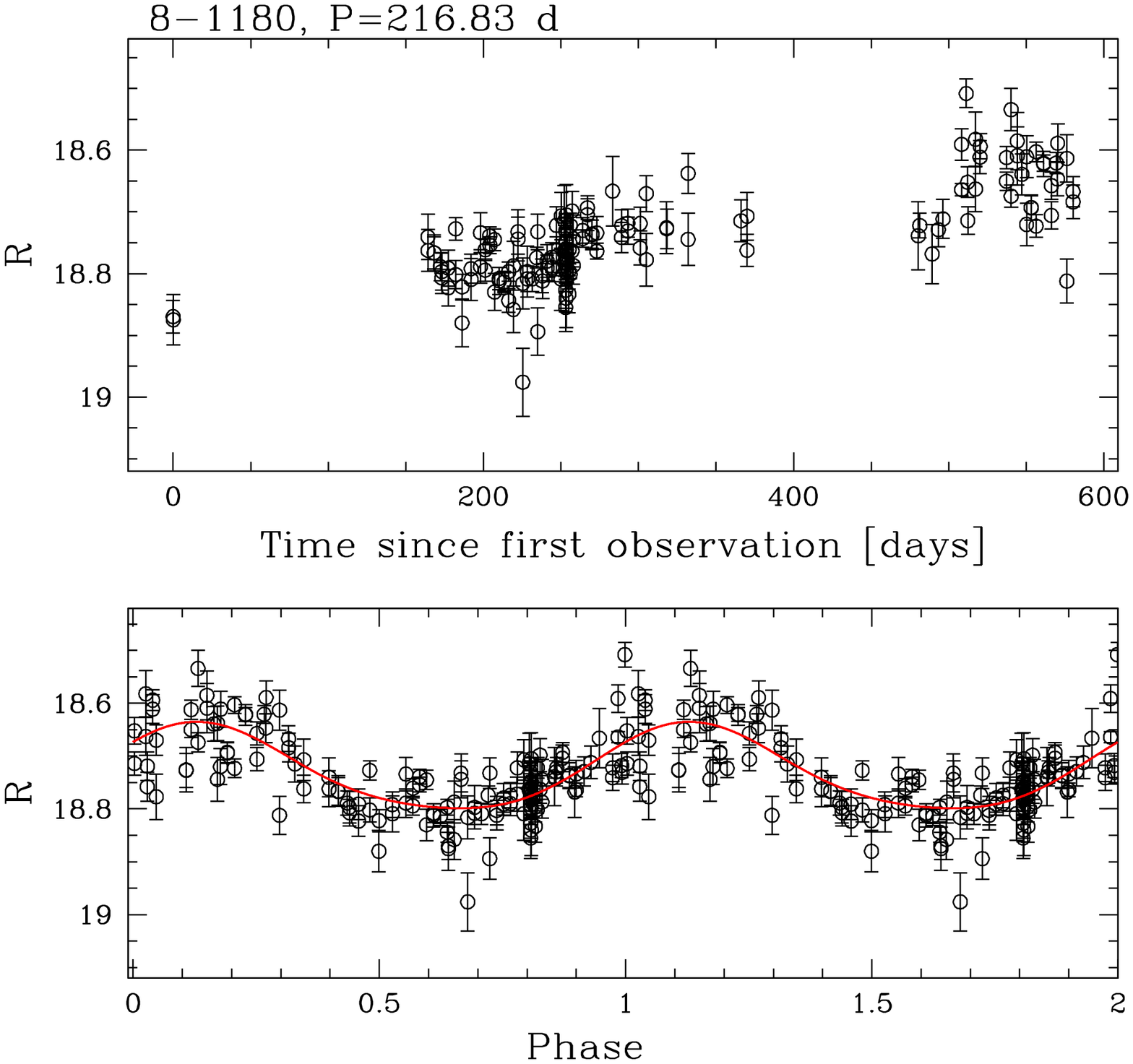} &
  \includegraphics[width=0.22\textwidth]{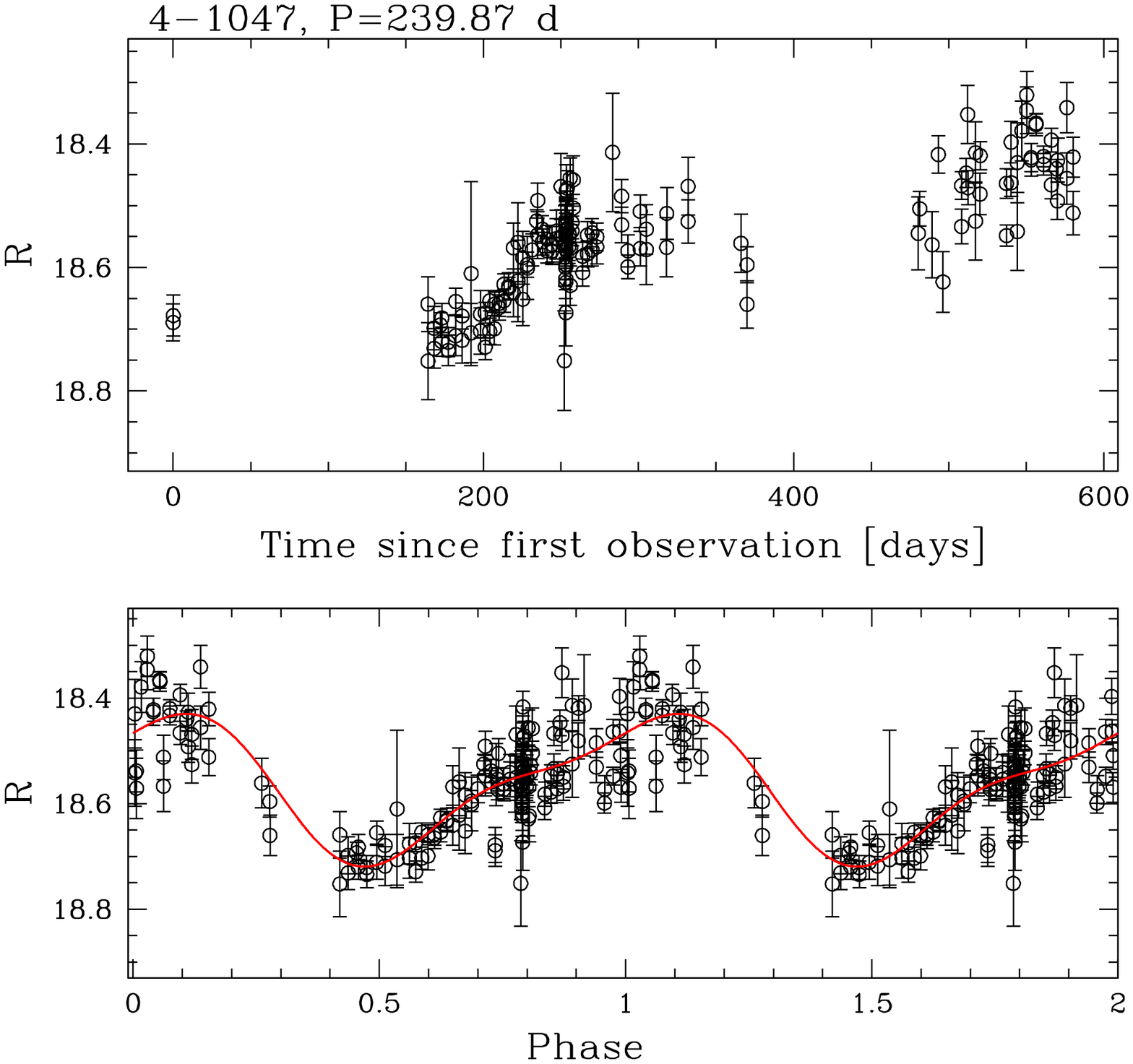} &
  \includegraphics[width=0.22\textwidth]{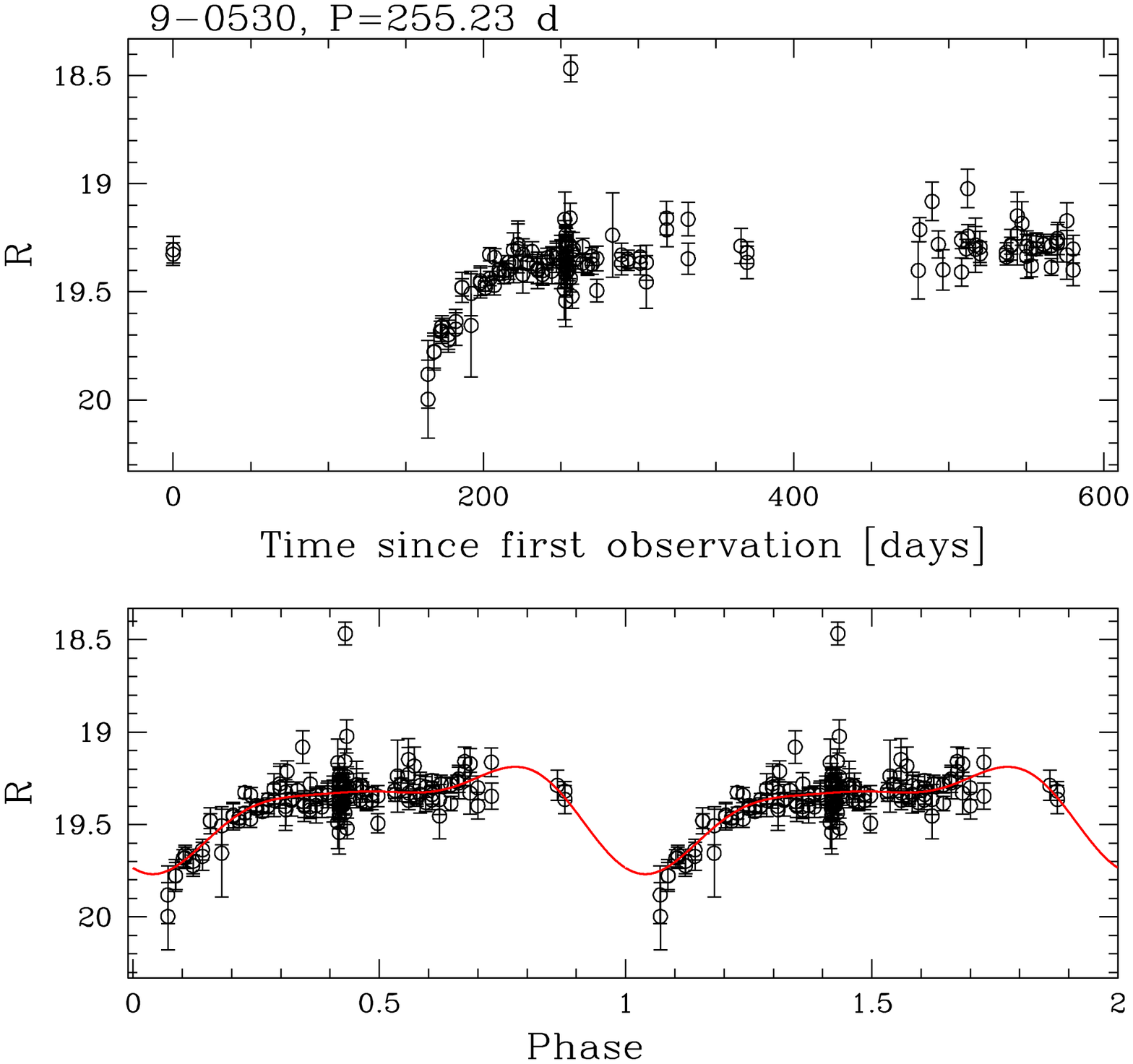} &
  \includegraphics[width=0.22\textwidth]{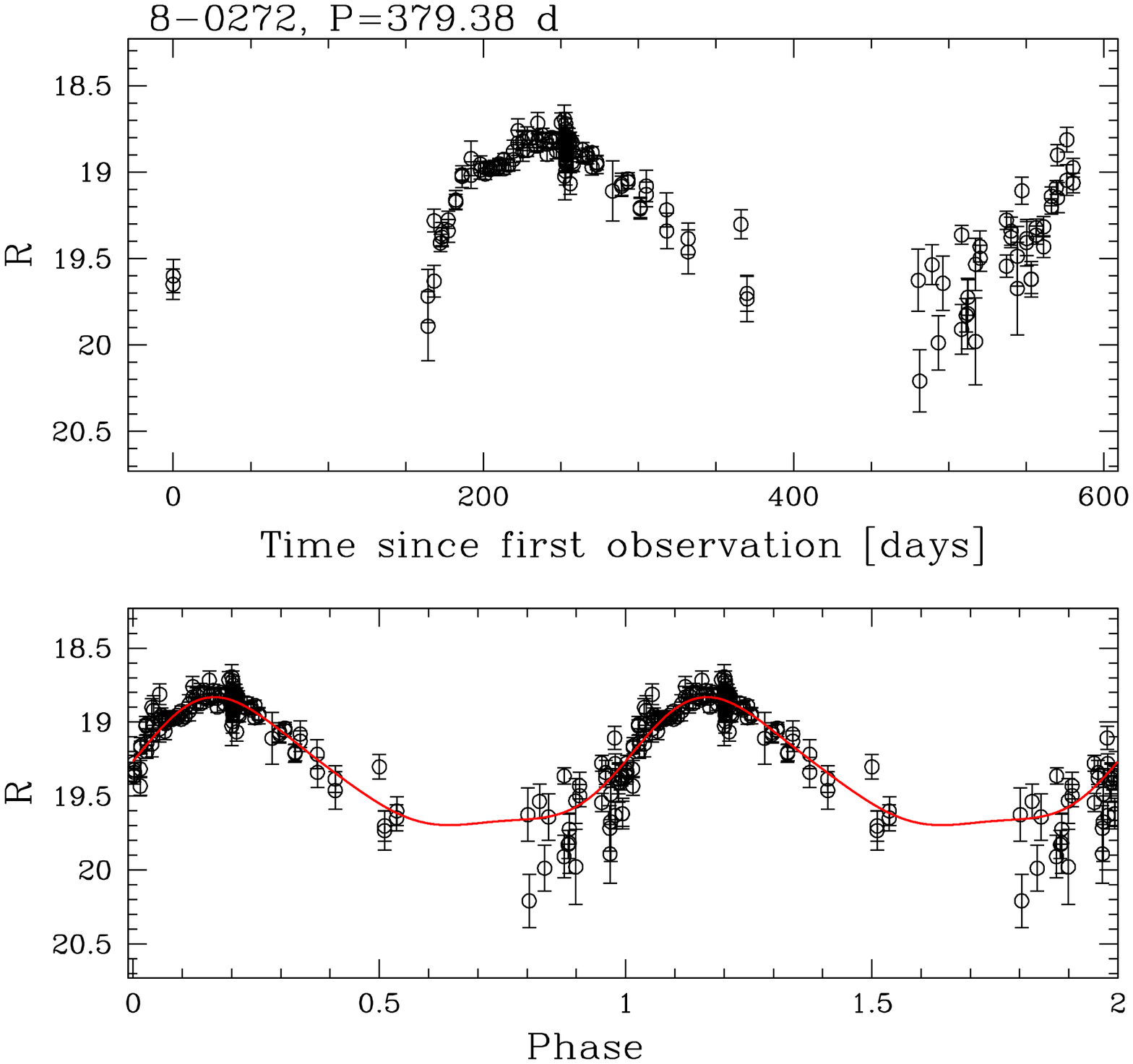} \\
   \end{array}$
  \caption{Light-curves of the 8 candidates that survived after the cuts (see Section~\ref{sec3}). The periods of these candidates are given in Table~\ref{tab2} and have been refined using the {\tt SigSpec} package. An unfolded light-curve (upper) and a folded light-curve (lower) is given for each candidate.}
  \label{fig.lc}
\end{figure*}

The next step was to build a reference image for the image subtraction from the 10 best-seeing images from all times. Since these images were taken at different seeing conditions, they were PSF-convolved to remove seeing variations and coadded to generate the reference image.  We then applied the image subtraction to all images. As part of this process, the reference image was convolved with a PSF kernel to match to each image. We subdivided each image into 36 pieces, with 6 sub-divisions along the $x$-axis and 6 sub-divisions along the $y$-axis, and constructed the kernel for each sub-division, in order to account for the spatially varying kernel.

After image subtraction, we built light-curves from the difference images. To search for ULPC candidates, we applied various selection criteria to the light-curves. These selection criteria are described as follows. 

\begin{itemize}
\item {\it Cut using magnitude range}: We utilized the P-L relation to derive the intrinsic brightnesses of Cepheids with periods longer than 80 days, and scaled their brightnesses to the distance of M31, in order to estimate their expected magnitudes. To extract sources that are in the magnitude range of ULPCs, we used {\tt SExtractor} to extract stars in the reference image and calibrated the magnitude zero point of the reference image relative to the Local Group Survey (LGS) catalog (\cite{massey06}). We then drew the light-curves on sources with magnitudes between 17.5 and 19.5, where the P-L relation predicts Cepheids with a period of 80 days to have a magnitude of around 18. We intentionally set the faint limit of the magnitude range to be 19.5, in order to incorporate the intergalactic and intrinsic extinction in the M31 line of sight. 

\item {\it Cut using power from analysis of variance (AoV)}: After identifying point sources within the adopted magnitude range, we applied the AoV algorithm (\cite{sc96}), which is ideal for period searches in unevenly sampled light-curves.  From inspection of some light-curves, we empirically required the power from AoV to fall between 0.5 and 0.9.

\item {\it Cut using period}: We utilized the AoV algorithm to find the preliminary period of each of the resolved sources within our adopted magnitude range. We allowed AoV to search for periodicity between 1 and 500 days. Because we are looking for ULPCs, we filtered out sources with periods of less than 80 days and larger than 300 days.  The surviving sources from this and the previous cuts then formed the subset of ULPC candidates subject to the next, final exploratory cut, which is further eye-inspection.

\item {\it Cut using eye inspection}: We still need to use human-eyes to identify the most likely ULPC candidates. In this step, we folded light-curves survived from the previous cuts, and inspected them by eye to see if they indicate prominent ULPC variabilities. 

\end{itemize}

The numbers of surviving candidates after applying the aforementioned cuts are summarized in Table~\ref{tab1}. Ultimately, 8 candidates were identified in our dataset. Figures~\ref{fig.lc} presents light-curves for these 8 candidates, while their locations relative to M31 can be seen in Figure~\ref{fig_m31}.

\section{Search Results}
\label{sec4}

\begin{table*}
\renewcommand{\arraystretch}{1.3}
\caption{Properties of our 8 candidates in M31.}
\label{tab2}
\centering
\begin{tabular}{ccccccccc}
\hline
Internal ID &
SIMBAD ID &
RA (J2000) & 
DEC (J2000) &
Period [day] &
$<R>$ &
$A_R$ &
$R_{21}$ &
$\phi_{21}$ \\
\hline
\multicolumn{9}{c}{ULPC candidates} \\
8-0326  & [MAP97] 55              & 00:42:31.838 & +41:29:10.67 &  74.942 & 18.156 & 0.166 & $0.261\pm0.026$ & $5.314\pm0.115$ \\
8-1498  & [MAP97] 59              & 00:42:53.551 & +41:03:12.41 &  86.210 & 18.565 & 0.166 & $0.254\pm0.031$ & $5.512\pm0.118$ \\
7-1326  & ---                     & 00:41:17.477 & +41:08:38.23 &  98.035 & 17.520 & 0.166 & $0.312\pm0.048$ & $6.230\pm0.167$ \\
\multicolumn{9}{c}{Other  Mira-like or ultra-long-period variables} \\
8-1176  & 2MASS J00432235+4110283 & 00:43:22.378 & +41:10:28.37 & 165.819 & 18.281 & 0.166 & $0.106\pm0.064$ & $3.565\pm0.492$ \\
8-1180  & 2MASS J00434756+4110284 & 00:43:47.554 & +41:10:28.36 & 216.830 & 18.738 & 0.166 & $0.141\pm0.037$ & $3.424\pm0.269$ \\
4-1047  & ---                     & 00:48:59.123 & +42:23:51.12 & 239.871 & 18.551 & 0.184 & $0.327\pm0.029$ & $1.228\pm0.076$ \\
9-0530  & ---                     & 00:45:45.305 & +41:28:51.10 & 255.232 & 19.353 & 0.166 & $0.691\pm0.051$ & $0.614\pm0.124$ \\
8-0272  & LGS J004324.24+413026.3 & 00:43:24.205 & +41:30:26.34 & 379.383 & 19.080 & 0.166 & $0.220\pm0.140$ & $3.364\pm0.273$ \\
\hline
\end{tabular}
\end{table*}

The properties of our 8 candidates are given in Table~\ref{tab2}. Throughout the paper, we use our internal ID when referring these candidates, where the first digit in this ID corresponds to the chip number as given in Table~\ref{tab1}, followed by 4 digits numbering from the DIA catalogs. We cross-matched these candidates with the SIMBAD\footnote{{\tt http://simbad.u-strasbg.fr/simbad/}} database, and list the looked-up IDs of the matched sources in the second column of Table~\ref{tab2}. The periods of these candidates have been refined using the {\tt SigSpec} package (\cite{reegan07}). Hence, the folded light-curves with these refined periods have smaller scatter, as shown in Figure~\ref{fig.lc}. The numerical values of the refined periods are reported in Table~\ref{tab2}.  

The $R$-band light-curve flux densities based on DIA have been converted into magnitudes, and calibrated to $R$-band magnitudes using the LGS catalog (\cite{massey06}). The mean $R$-band magnitudes ($<R>$) and the foreground $R$-band extinction ($A_R$) for the 8 candidates, estimated based on the \cite{schlegel98} extinction map, are listed in the sixth and seventh columns of Table~\ref{tab2}, respectively. The folded $R$-band light-curves were then fitted with a low-order Fourier expansion (for example, see \cite{simon81}) in the form of:

\begin{eqnarray}
R(\phi) & = & A_0 + \sum_{i=1}^{i=2\ \mathrm{or }\ 3} A_i \cos(2\pi i\phi + \phi_i), \nonumber
\end{eqnarray}

\noindent where $\phi$ is the folded phase. The fitted light-curves are shown as (red) curves in the folded light-curve scatter plots of Figure~\ref{fig.lc}. The Fourier parameters, $R_{21}=A_2/A_1$ and $\phi_{21}=\phi_2-2\phi_1$, and their associated errors (\cite{petersen86}) for the ULPC candidates are given in the last two columns of Table~\ref{tab2}. 

\begin{figure}
\centering
\includegraphics[width=3.2in]{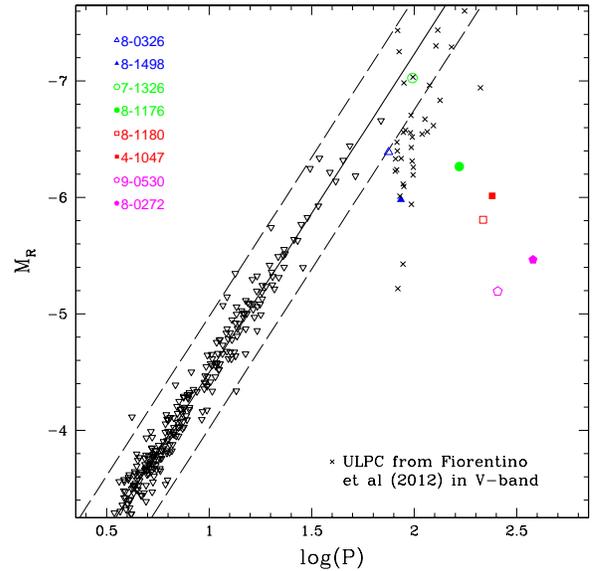}
\caption{The P-L relation for our 8 candidates (various non-black, colored symbols) as compared to the Galactic Cepheids (upside-down open triangles) in the $R$-band. The mean $R$-band magnitudes for the 8 candidates have been corrected for extinction as listed in Table~\ref{tab2}, and a distance modulus of $24.38$ from \cite{riess12} was adopted. The data for the Galactic Cepheids were taken from \cite{ngeow12}. The thick line represents the fitted P-L relation for Galactic Cepheids, again taken from \cite{ngeow12}. The dashed lines are the $\pm 3\sigma_{{\mathrm t}}$ of the P-L relation, where $\sigma_{{\mathrm t}}$ includes the dispersion of the $R$-band P-L relation, and the errors (both random and systematic error) in distance modulus. The ULPCs from \cite{fiorentino12} in the $V$-band (diagonal crosses) are included for comparison. Note that the large scatter of these ULPCs include errors from photometry, distance, extinction, and the intrinsic dispersion of the P-L relation for ULPCs.}
\label{fig.pl}
\end{figure}

\begin{figure}
\centering
\includegraphics[width=2.8in]{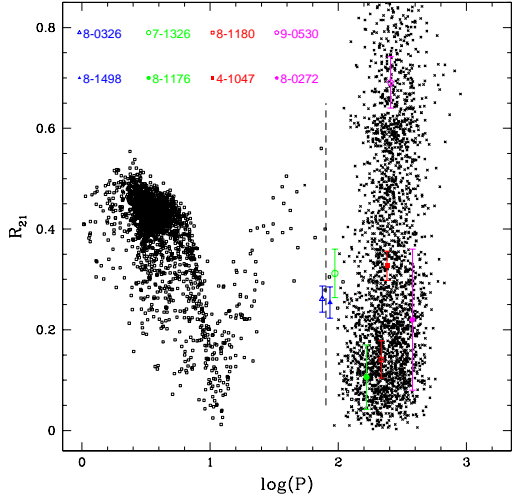}
\caption{Comparison of the $R_{21}$ parameters for our 8 candidates (various non-black, colored symbols) to Cepheids (black open squares) and Mira variables (diagonal crosses). The $R_{21}$ values for both Cepheids and Mira variables were adopted from \cite{deb09}. The dashed vertical line indicates the period of $80$ days.}
\label{fig.deb}
\end{figure}

\begin{figure}
\centering
\includegraphics[width=2.8in]{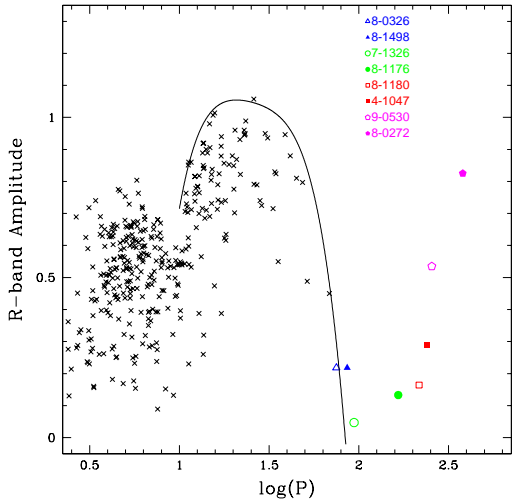}
\caption{Comparison of the $R$-band amplitudes for Galactic fundamental-mode Cepheids (diagonal crosses, adopted from \cite{klagyivik09}) and our 8 candidates (various non-black, colored symbols; derived from the fitted light-curves with Fourier expansion). The curve represents the upper envelope for Galactic Cepheids with periods greater than $10$ days, derived using the {\tt boundfit} code (\cite{cardiel09}) with a simple polynomial expansion, and extrapolated to $\log(P)\sim1.9$.}
\label{fig.amp}
\end{figure}

In Figures~\ref{fig.pl},~\ref{fig.deb} and~\ref{fig.amp}, the $R$-band absolute magnitudes, $R_{21}$ values and $R$-band amplitudes for the 8 candidates were compared to the classical Cepheids in the P-L plane, the period-$R_{21}$ plane and the period-amplitude plane, respectively. Assuming $M_V\sim M_R$, the ULPCs listed in \cite{fiorentino12} can be included in Figure~\ref{fig.pl} (as most of the ULPCs have $V$- and $I$-band data, but not $R$-band data). In Figure~\ref{fig.deb}, we also include the $R_{21}$ values for Mira variables adopted from \cite{deb09}. Figures 6--8 reveal that the 3 candidates with shorter periods (8-0326, 8-1498 \& 7-1326) could be genuine ULPCs (see further discussion in the following subsections). The other 5 longer-period candidates are probably Mira-like or ultra-long-period variables, and will not be discussed any further in this paper.   

\subsection{Internal ID 8-0326}

A Cepheid with a period of $75$ days in M31 has been reported by \cite{ivanov85}, and confirmed by \cite{ivanov86} (with period of $75.487$ days). We have recovered this Cepheid from PTF data, with a slightly shorter period (see Table~\ref{tab2}).  The Cepheid nature of this candidate can also be verified from its location in P-L relation (Figure~\ref{fig.pl}) and Fourier parameter (Figure~\ref{fig.deb}). The amplitude of this candidate lies along the envelope of the period-amplitude diagram extrapolated from long-period Cepheids (solid curve in Figure~\ref{fig.amp}). The SIMBAD database has identified this Cepheid to be [MAP97] 55 from \cite{magnier97}. However the period reported in their catalog is $40\pm8$ days, about half of the period found in this work. The $\approx75$-day period of this Cepheid is slightly shorter than the cut-off period of $80$ days between classical Cepheids and ULPCs defined in \cite{bird09}. This case is an example of a Cepheid that is on the borderline of the commonly accepted cut-off period for defining classical Cepheids vs. ULPCs.

\subsection{Internal ID 8-1498}

The SIMBAD database identified this candidate as a Cepheid variable from the \cite{magnier97} catalog. The period of $50\pm10$ days given in that catalog is 42\% shorter than the period of 86.2 days found in this work (see Table~\ref{tab2}). The light-curve shape (Figure~\ref{fig.lc}) and $R_{21}$ parameter value (Figure~\ref{fig.deb}) for this candidate are representative of typical Cepheid-like and ULPC-like light-curves. The amplitude of the $R$-band light-curve for this candidate, which is similar to the previous candidate, also lies near the curve plotted in the period-amplitude diagram of Figure~\ref{fig.amp}, which represents the upper envelope for Galactic Cepheids with periods greater than $10$ days. The $R$-band mean magnitude for this candidate is fainter as compared to candidate \#8-0326 at a similar period (see Figure~\ref{fig.pl}), and appears to be an outlier. However, one should keep in mind that the dispersions of P-L relations for ULPCs are $\sim1.6\times$ larger than the Cepheid counterparts \cite{bird09}. Inspecting Figure~\ref{fig_m31} suggests that this candidate may be located at the back, or far-side, of M31, which could suffer from a depth-effect and/or higher extinction.

\subsection{Internal ID 7-1326}

The $R$-band absolute magnitude for this candidate falls on the P-L relation defined by the Galactic Cepheids (Figure~\ref{fig.pl}). The $R_{21}$ value for this candidate is also consistent with ULPCs (Figure~\ref{fig.deb}). Interestingly, the amplitude of the candidate is quite low as compared to other candidates. This candidate may be an ULPC evolving off or near the instability strip of ULPCs. No counterpart was found in SIMBAD database for this candidate.

\section{Discussion and Conclusion}

We have presented a comprehensive search for M31 ULPCs using PTF data that was acquired over a time period of approximately 2 years. Benefiting from the large FOV, high cadence, and long time-baseline of the PTF observing strategy, we have been able to surpass previous M31 monitoring campaigns and trace ULPCs even out to the halo of M31. We have utilized a difference image method to reveal source variability in crowded stellar fields, and focused on point sources in the magnitude range of ULPCs inferred by P-L relation. 

\begin{figure}
\centering
\includegraphics[width=2.5in]{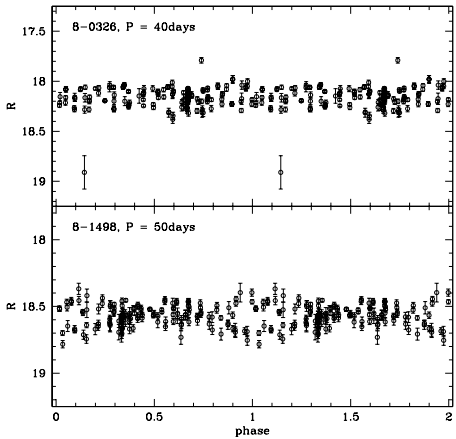}
\caption{Folded light curves for ULPC candidates 8-0326 (upper panel) and 8-1498 (lower panel) if using the published periods given in \cite{magnier97}.}
\label{fig.wrongp}
\end{figure}

We have selected 8 candidates, and performed Fourier-decomposition to derive their mean magnitude, amplitude, $R_{21}$, and $\phi_{21}$ from $R$-band light-curves. Among the 8 candidates, we characterized 3 of the most promising ULPC candidates by their location on the P-L relation, $R_{21}$ vs.\ period, and $R$-band amplitude vs.\ period parameter plane. The mean $R$-band absolute magnitude for these 3 candidates, calibrated using the distance modulus of M31 from \cite{riess12}, is $M_R=-6.47$mag (with a dispersion of $0.53$mag). Two of the candidates (8-0326 \& 8-1498) have been reported before (\cite{ivanov86,magnier97}). However, periods reported from \cite{magnier97} for these 2 candidates do not agree with the periods we found here (Table~\ref{tab2}). Inspecting the light curves presented in \cite{magnier97} reveal that their light curves do not represent the full pulsation cycle of these 2 candidates. Figure~\ref{fig.wrongp} presents the folded light curves using our data but with \cite{magnier97}'s periods, none of them exhibit Cepheid-like or ULPC-like light curves. The third candidate, 7-1326, is a new discovery, probably due to it's low amplitude. Confirmation of the ULPC nature of these 3 candidates has to await the availability of $V$ and $I$ band light-curves, from which the accurate mean $VI$-band magnitudes can be deduced. The mean $VI$-band magnitudes are important in constructing the color-magnitude diagram and the extinction-free Wesenheit function, defined as $W=I-1.55(V-I)$, for these candidates, so that they can be compared to the known ULPCs as given in \cite{fiorentino12}. The period-Wesenheit (PW) plot for these candidates is particularly important because the dispersion of the PW relation for ULPCs is much smaller than that of $VI$-band P-L relations (\cite{bird09}).

ULPCs are evolved stars with mass between $\sim15M_{\odot}$ to $\sim20M_{\odot}$ (\cite{bird09,fiorentino12}); hence, the number of ULPCs in a given galaxy will not be many. Therefore, it is not a surprise that only a few ULPC candidates were found in this work (for comparison, only two ULPCs were found in M81; see \cite{fiorentino12}). Furthermore, the ULPC candidates found in this work are distributed farther away from the M31 center, as shown in Figure~\ref{fig_m31}. This can be partially attributed to high-level background from M31 itself and shallower photometry from the relatively short exposure time of PTF data (60~s per image). The high-level background of M31 also brings up concern about the issue of source blending and/or crowding. This is because the M31 background consists of numerous stars, and at the distance of M31, each pixel can contain many more than a single star (see e.g., \cite{riffeser06}). The crowding issue is severer for P48, because the pixel size is 1.01~arcsec and the average seeing is about 2.5~arcsec (see Figure~\ref{fig.psf}). Nevertheless, the crowding issue can be solved by adopting an image-subtraction algorithm, as we did in Section~\ref{sec3}. The issue of blending is hard to tackle with since the pixel size and the site seeing of P48 are huge compared to other M31 monitoring campaigns. To have a precise estimation of the influence of blending, one can follow the steps of \cite{chaves12}, where they compared {\it HST} images and ground-based observations to infer the blending effects in M33. However, this is beyond the scope of the aim of this work. In addition, the ULPCs found in this work are rather bright compared to normal Cepheids, which will be less affected by blending from near-by stars.

\section*{Acknowledgment}

This work was partly supported by the NSC grants: NSC101-2119-M-008-007-MY3 and NSC101-2628-M-008-003. This research has made use of the SIMBAD database, operated at CDS, Strasbourg, France. We also thank S. Deb for sharing the $R_{21}$ data from his paper.

% that's all folks

\begin{thebibliography}{1}

\bibitem{bird09} Bird, J.~C., Stanek, K.~Z., \& Prieto, J.~L.\ 2009, The Astrophysical Journal, Vol. 695, pg. 874 

\bibitem{fiorentino12} Fiorentino, G., Clementini, G. Marconi, M. ,et al.\ 2012, Astrophysics and Space Science, Vol. 341, pg. 143

\bibitem{vilardell10} Vilardell, F., Ribas, I., Jordi, C., Fitzpatrick, E.~L., \& Guinan, E.~F.\ 2010, Astronomy and Astrophysics, Vol. 509, pg. A70 

\bibitem{rau09} Rau, A., Kulkarni, S.~R., Law, N.~M., et al.\ 2009, Publications of the Astronomical Society of the Pacific, Vol. 121, pg. 1334 

\bibitem{law09} Law, N.~M., Kulkarni, S.~R., Dekany, R.~G., et al.\ 2009, Publications of the Astronomical Society of the Pacific, Vol. 121, pg. 1395 

\bibitem{grillmair10} Grillmair, C.~J., Laher, R., Surace, J., et al.\ 2010, Astronomical Data Analysis Software and Systems XIX, edited by Y. Mizumoto, K.-I. Morita \& M. Ohishi, ASP Conference Series, Vol. 434, pg. 28

\bibitem{ofek12} Ofek, E.~O., Laher, R., Law, N., et al.\ 2012, Publications of the Astronomical Society of the Pacific, Vol. 124, pg. 62 

\bibitem{alard98} Alard, C., \& Lupton, R.~H.\ 1998, The Astrophysical Journal, Vol. 503, pg. 325 

\bibitem{bertin96} Bertin, E., \& Arnouts, S.\ 1996, Astronomy and Astrophysics Supplement, Vol. 117, pg. 393 

\bibitem{massey06} Massey, P., Olsen, K.~A.~G., Hodge, P.~W., et al.\ 2006, The Astronomical Journal, Vol. 131, pg. 2478 

\bibitem{sc96} Schwarzenberg-Czerny, A.\ 1996, The Astrophysical Journal Letters, Vol. 460, pg. L107 

\bibitem{reegan07} Reegen, P.\ 2007, Astronomy and Astrophysics, Vol. 467, pg. 1353 

\bibitem{schlegel98} Schlegel, D.~J., Finkbeiner, D.~P., \& Davis, M.\ 1998, The Astrophysical Journal, Vol. 500, pg. 525

\bibitem{simon81} Simon, N.~R., \& Lee, A.~S.\ 1981, The Astrophysical Journal, Vol. 248, pg. 291 

\bibitem{petersen86} Petersen, J.~O.\ 1986, Astronomy and Astrophysics, Vol. 170, pg. 59 

%\bibitem{petersen94} Petersen, J.~O.\ 1994, Astronomy and Astrophysics, Vol. 105, pg. 145 

\bibitem{riess12} Riess, A.~G., Fliri, J., \& Valls-Gabaud, D.\ 2012, The Astrophysical Journal, Vol. 745, pg. 156 

\bibitem{ngeow12} Ngeow, C.-C.\ 2012, The Astrophysical Journal, Vol. 747, pg. 50 

\bibitem{deb09} Deb, S., \& Singh, H.~P.\ 2009, Astronomy and Astrophysics, Vol. 507, pg. 1729 

\bibitem{klagyivik09} Klagyivik, P., \& Szabados, L.\ 2009, Astronomy and Astrophysics, Vol. 504, pg. 959 

\bibitem{cardiel09} Cardiel, N.\ 2009, Monthly Notices of the Royal Astronomical Society, Vol. 396, pg. 680 

\bibitem{ivanov85} Ivanov, G.~R.\ 1985, Astrophysics and Space Science, Vol. 115, pg. 409 

\bibitem{ivanov86} Ivanov, G.~R., \& Sharov, A.~S.\ 1986, Astrophysics and Space Science, Vol. 124, pg. 329 

\bibitem{magnier97} Magnier, E.~A., Augusteijn, T., Prins, S., van Paradijs, J., \& Lewin, W.~H.~G.\ 1997, Astronomy and Astrophysics Supplement, Vol. 126, pg. 401 

\bibitem{riffeser06} Riffeser, A., Fliri, J., Seitz, S., \& Bender, R.\ 2006, The Astrophysical Journal Supplement, Vol. 163, pg. 225

\bibitem{chaves12} Chavez, J.~M., Macri, L.~M., \& Pellerin, A.\ 2012, The Astronomical Journal, Vol. 144, pg. 113 


\end{thebibliography}
\end{document}